\documentclass[sigconf]{acmart}

\AtBeginDocument{%
  \providecommand\BibTeX{{%
    \normalfont B\kern-0.5em{\scshape i\kern-0.25em b}\kern-0.8em\TeX}}}

\usepackage{multirow}
\usepackage{ragged2e}
\usepackage{tabularx}

\usepackage{color}
\usepackage{colortbl}

\usepackage{todonotes}
\let\xtodo\todo
\renewcommand{\todo}[1]{\xtodo[inline]{#1}}

\copyrightyear{2021}
\acmYear{2021}
\setcopyright{ccbysa}
\acmConference[Emotion Workshop '22]{The Future of Emotion in Human-Computer Interaction}{April 13, 2022}{New Orleans, USA}
\acmBooktitle{Emotion Workshop '22 – The Future of Emotion in Human-Computer Interaction, April 13, 2022, New Orleans, USA}
\acmPrice{15.00}

\newcommand{\yes}{\cellcolor[HTML]{B7E1CD}Yes}
\newcommand{\no}{\cellcolor[HTML]{F4CCCC}No}

\begin{document}

\title{Current Challenges of Using Wearable Devices for Online Emotion Sensing}


\author{Weiwei Jiang}
\orcid{0000-0003-4413-2497}
\affiliation{
    \institution{The University of Melbourne}
    \city{Melbourne}
    \postcode{3010}
    \country{Australia}
}
\email{weiwei.jiang@student.unimelb.edu.au}

\author{Kangning Yang}
\orcid{0000-0002-7106-0022}
\affiliation{
    \institution{The University of Melbourne}
    \city{Melbourne}
    \postcode{3010}
    \country{Australia}
}
\email{kangning@student.unimelb.edu.au}

\author{Maximiliane Windl}
\orcid{0000-0002-9743-3819}
\affiliation{%
  \institution{LMU Munich}
  \city{Munich}
  \postcode{80337}
  \country{Germany}
}
\email{maximiliane.windl@ifi.lmu.de}

\author{Francesco Chiossi}
 \orcid{0000-0003-2987-7634}
\affiliation{%
  \institution{LMU Munich}
  \city{Munich}
  \postcode{80337}
  \country{Germany}
}
\email{francesco.chiossi@um.ifi.lmu.de}

\author{Benjamin Tag}
\orcid{0000-0002-7831-2632}
\affiliation{
    \institution{The University of Melbourne}
    \city{Melbourne}
    \postcode{3010}
    \country{Australia}
}
\email{benjamin.tag@unimelb.edu.au}

\author{Sven Mayer}
\orcid{0000-0001-5462-8782}
\affiliation{%
  \institution{LMU Munich}
  \city{Munich}
  \postcode{80337}
  \country{Germany}
}
\email{info@sven-mayer.com}

\author{Zhanna Sarsenbayeva}
\orcid{0000-0002-1247-6036}
\affiliation{
    \institution{The University of Melbourne}
    \city{Melbourne}
    \postcode{3010}
    \country{Australia}
}
\email{zhanna.sarsenbayeva@unimelb.edu.au}

\renewcommand{\shortauthors}{Jiang, et al.}

\begin{abstract}
A growing number of wearable devices is becoming increasingly non-invasive, readily available, and versatile for measuring different physiological signals. This renders them ideal for inferring the emotional states of their users. Despite the success of wearable devices in recent emotion studies, there are still several challenges to be addressed. In this position paper, we compare currently available wearables that can be used for emotion-sensing and identify the challenges and opportunities for future researchers. Our investigation opens the discussion of what is missing for in-the-wild for emotion-sensing studies.
\end{abstract}



\maketitle

\section{Introduction}
Emotions are an integral part of human life and influence how we behave and think. Emotion detection research has gained significant attention across different scientific disciplines, including Computer Science and Human-Computer Interaction (HCI) (e.g., mobile interaction under stress~\cite{sarsenbayeva2019measuring}, correlation between emotion and smartphone usage~\cite{sarsenbayeva2020does}), both from the perspective of conducting research and building emotion-sensing applications. Among a variety of different approaches to emotion-sensing~\cite{tag2021retrospective}, the most widely used methods for measuring emotions are self-reports of subjective experience~\cite{sas2010investigating}. These require participants to be aware of their experienced emotions and reflect accurately on their phenomenal awareness through rating scales~\cite{sas2010investigating}. While self-reports are considered the gold standard for ground truth data collection~\cite{tag2021retrospective}, self-reports are prone to inaccuracies; thus, keeping the questions of how to measure human emotions objectively unanswered~\cite{gross2003individual}.

Researchers have attempted to use physiological measures~\cite{shu2019emotion}, such as electroencephalography (EEG), electrocardiogram (ECG), body temperature, and electrodermal activity (EDA) to objectively quantify human cognitive state. Alternatively, a number of approaches are vision- or audio-based methods (such as facial expression and voice). Here, physiological measures help to understand the emotional state as they represent involuntary responses that cannot be easily controlled or hidden~\cite{shu2018review}. However, due to the required constant monitoring of physiological measures privacy is at risk~\cite{shu2019emotion,prange2021investigating}. 
In particular, affective computing researchers nowadays rely on wearable devices that are  non-invasive, readily available, and versatile in collecting various types of physiological data (cf. \cite{tag2021making,chiossi2022designing}). Measuring physiological signals is the first step in extracting the user's emotions in real time. The next challenge of how to analyze and then interpret this data stream.

In this position paper, we identify the challenges and future directions of using wearable sensors for emotion sensing. We first summarize existing wearable sensors that have recently been widely used for emotion-sensing by researchers. Then, we highlight the future directions that promise to benefit wearable emotion sensing studies.

\begin{table*}[t]
\small
\caption{Sampling wearable products with physiological sensors that can be used for sensing emotions}
\label{tab:emotion_sensors}
\begin{tabularx}{\linewidth}{Xl*{9}{c}}
\toprule
 \multirow{2}{*}{\textbf{Product}} & \multirow{2}{*}{\textbf{Type}} & \multicolumn{9}{c}{\textbf{Available Sensors}} \\
 \cline{3-11}
   &
   & \it IMU & \it BP Cuff & \it ECG & \it EDA & \it EEG & \it EMG & \it PPG & \it Temp. & \it SpO2 \\
  \midrule
Apple Watch 7              & Wrist      & \yes & \no  & \yes & \no  & \no  & \no  & \no  & \no  & \yes \\
Bitalino PsychoBIT BT      & Whole body & \no  & \no  & \yes & \yes & \no  & \no  & \no  & \no  & \no  \\
Brain Vision EcgMove 4     & Chest      & \yes & \no  & \yes & \no  & \no  & \no  & \no  & \yes & \no  \\
Brain Vision EdaMove 4     & Wrist      & \yes & \no  & \no  & \yes & \no  & \no  & \no  & \no  & \no  \\
Brain Vision LiveAmp 64    & Head       & \yes & \no  & \no  & \no  & \yes & \no  & \no  & \no  & \no  \\
DSI VR300                  & Head       & \no  & \no  & \no  & \no  & \yes & \no  & \no  & \no  & \no  \\
Emotiv Epoc+               & Head       & \no  & \no  & \no  & \no  & \yes & \no  & \no  & \no  & \no  \\
Empatica E3/E4/EmbracePlus & Wrist      & \yes & \no  & \no  & \yes & \no  & \no  & \yes & \yes & \no  \\
Fitbit Sense               & Wrist      & \yes & \no  & \yes & \yes & \no  & \no  & \yes & \yes & \yes \\
M Brain Trin Smarting Mobi & Head       & \no  & \no  & \no  & \no  & \yes & \no  & \no  & \no  & \no  \\
Microsoft Band 2           & Wrist      & \no  & \no  & \no  & \no  & \no  & \no  & \yes & \yes & \no  \\
Myo Armband                & Arm        & \no  & \no  & \no  & \no  & \no  & \yes & \no  & \no  & \no  \\
NeuroSky MindWave Mobile 2 & Head       & \no  & \no  & \no  & \no  & \yes & \no  & \no  & \no  & \no  \\
OpenBCI Biosensing         & Head       & \no  & \no  & \yes & \no  & \yes & \yes & \no  & \no  & \no  \\
Polar H10                  & Chest      & \no  & \no  & \yes & \no  & \no  & \no  & \no  & \no  & \no  \\
Vivalink Remote Monitoring & Whole body & \no  & \yes & \yes & \no  & \no  & \no  & \no  & \yes & \yes \\
  \toprule
\end{tabularx}
\end{table*}

\section{Commercially Available Emotion Sensors}
Despite the advantages of using wearable sensors for physiological measurements, it is not trivial to develop a reliable wearable device for such purposes~\cite{seneviratne2017survey}. Hence, many researchers rely on (validated) existing products, especially for studying emotion. As a reference, we show a sample of frequently used wearable devices with physiological sensors that are commercially available in \autoref{tab:emotion_sensors}. Here, there are a variety of measurements: EEG as a measure of brain activity on the scalp, EDA a measure of skin conductance, Electromyography (EMG) a measure of skeletal muscle activity, and ECG a measure of heart activity.

In particular for sensing emotions, physiological measures offer the great advantage of capturing the data in real-time and continuously. This is in stark contrast to more subjective methods such as the Experience Sampling Method (ESM)~\cite{vanberkel2017experience}. For instance, EEG responses can be correlated to emotional responses~\cite{shu2019emotion,harmon2003clarifying}. Other physiological signals may also be affected by emotions such as EDA~\cite{Babaei2021}, EMG, and ECG. 

Beyond these biosignals, researchers also measure other biomarkers that may be correlated to emotions. Unlike biosignal sensors, biomarker sensors measure the desired signals indirectly, as some biomarker may not be directly measured~\cite{nimse2016biomarker}. For instance, Photoplethysmography sensors (PPG) are widely used in wrist-worn devices for measuring blood volume changes to infer HRV. In particular for emotion sensing, HRV is considered more useful for stress monitoring than simple heart rate monitoring~\cite{sarsenbayeva2019measuring}. 

In addition to sensor availability, existing devices can be worn on different body locations, depending on the signals to be measured. For instance, EEG senors that measure brain activities must be attached to the head at very specific locations, while ECG sensors that measure heart activities must be attached to the chest or limbs, thus, limiting mobility of the wearer. Other sensors with less constrains such as PPG and EDA can be rather flexibly placed around the wrist or fingers. Together with this, wearable devices measuring ECG and EDA showed equivalent accuracy when compared to gold-standard equipment in different conditions~\cite{menghini2019stressing}. Consequently, to comprehensively sense physiological signals that are related to emotions, researchers are required to understand the plethora of sensors available and combinations possible to obtain a comprehensive and continuous data stream.

\section{Challenges and Opportunities}
As we can see in \autoref{tab:emotion_sensors}, most wearable devices are only capable of sensing specific physiological signals. Nevertheless, recent studies show that using multimodal approaches to emotion sensing can increase the detection performance~\cite{dmello2015review}. The reason for this is that people are unlikely to experience a single pure form of an emotion that only stimulates one specific physiological system~\cite{shu2019emotion}. Therefore, researchers need to address the challenge of using multiple devices for acquiring multimodal physiological signals, including deployment, signal synchronization, data transformation and other signal processing procedures~\cite{shu2019emotion}.

Even though an increasing number of sensors is becoming available in emerging wearables, only a few of them are dedicated to emotion sensing. Albeit some development kit (e.g., Bitalino) provides a modularized design for measuring multiple physiological data, it is used for prototyping and may not be ready for long-term in-the-wild studies~\cite{larradet2020toward}.

To this end, we still lack a dedicated wearable device that enables multimodal physiological sensing to help emotion researchers improve data collection comprehensively. A major challenge of developing such a device is that some physiological sensors are required to be attached to designated human body locations as mentioned above. To mitigate these issues, future deployments should consider indirect measurement methods. For example, recently it has been shown that respiration can be measured using WiFi signal~\cite{khan2017deep}. Furthermore, researchers demonstrated that some emotions can be recognized using wireless signals such as EQ-Radio~\cite{zhao2016emotion} or RFID signals~\cite{xu2020emotion}. In addition, as existing physiological sensors are designed for generic health care purpose, their technical specifications (e.g., granularity and sampling rate) may be either insufficient or overshooting the requirements of emotion sensing. To summarize, we argue that a wearable device with multimodal physiological sensing would benefit the data collection and, hence, improve the development of emotion research significantly. 

\section{Summary}
In this position paper, we compared currently available wearables that can be used for emotion-sensing and opened the discussion of what is missing for in-the-wild deployment for emotion-sensing studies. In particular, we highlighted that a multimodal wearable device dedicated for emotion sensing is highly desired that can greatly benefit to practical emotion-sensing studies and applications.

\begin{acks}
This work is partially funded by the Australia-Germany Joint Research Co-operation Scheme. Zhanna Sarsenbayeva is supported by a Doreen Thomas Postdoctoral Fellowship. Weiwei Jiang and Kangning Yang are supported by Melbourne Research Scholarship.
Francesco Chiossi is funded by the Deutsche Forschungsgemeinschaft (DFG, German Research Foundation) – Project-ID 251654672 – TRR 161.
Sven Mayer and Maximiliane Windl were funded by the German Federal Ministry of Education and Research (Bundesministeriums f\"{ü}r Bildung und Forschung) for this paper. However, the responsibility for the content of this publication lies with the authors.
\end{acks}

\balance
\bibliographystyle{ACM-Reference-Format}
\bibliography{main}

\end{document}